\documentclass[twocolumn,showpacs,preprintnumbers]{revtex4}
\usepackage{amsfonts}
\usepackage{amsmath}
\usepackage{amssymb}
\usepackage{graphicx}

\begin{document}

\title{Stationary light in cold atomic gases}
\author{Gor Nikoghosyan$^{1,2}$ and Michael Fleischhauer$^{1}$}
\affiliation{$^{1}$Fachbereich Physik and research center OPTIMAS, Universit\"{a}t
Kaiserslautern, Erwin-Schr\"{o}dinger-Strasse, D-67663 Kaiserslautern,
Germany\\
$^{2}$Institute of Physical Research, 378410, Ashtarak-2, Armenia}

\begin{abstract}
We discuss stationary light created by a pair of counter-propagating control
fields in $\Lambda$-type atomic gases with electromagnetically induced transparency
for the case of negligible Doppler broadening. In this case the secular
approximation used in the discussion of stationary light in hot vapors is no longer valid. We discuss
the quality of the effective light-trapping system and show that in contrast to previous
claims it is finite even for vanishing ground-state dephasing.
The dynamics of the photon loss is in general non exponential and can be
faster or slower than in hot gases.
\end{abstract}

\maketitle

%%%%%%%%%%%%%%%%%%%%%%%%%%%%%%%%%%%%%%%%%%%%%%%%%%%%%%%%%%%%%%%%%%%%%%%%%%%%%%%%

%%%%%%%%%%%%%%%%%%%%%%%%%%%%%%%%%%%%%%%%%%%%%%%%%%%%%%%%%%%%%%%%%%%%%%%%%%%%%%%%

%%%%%%%%%%%%%%%%%%%%%%%%%%%%%%%%%%%%%%%%%%%%%%%%%%%%%%%%%%%%%%%%%%%%%%%%%%%%%%%%

%%%%%%%%%%%%%%%%%%%%%%%%%%%%%%%%%%%%%%%%%%%%%%%%%%%%%%%%%%%%%%%%%%%%%%%%%%%%%%%%

%%%%%%%%%%%%%%%%%%%%%%%%%%%%%%%%%%%%%%%%%%%%%%%%%%%%%%%%%%%%%%%%%%%%%%%%%%%%%%%%
\section{Introduction}
%%%%%%%%%%%%%%%%%%%%%%%%%%%%%%%%%%%%%%%%%%%%%%%%%%%%%%%%%%%%%%%%%%%%%%%%%%%%%%%%
Strong coupling between light and matter is of large interest in many fields of
physics. It is of particular importance in quantum information and
quantum-optical realizations of strongly interacting many-body systems.
The interaction strength between single photons and
quantum dipole oscillators is determined by the value of the electromagnetic
field at the position of the oscillator and thus by the spatial
confinement of the photons. This has lead to the development
of cavity quantum electrodynamics where strong confinement and thus
strong coupling is achieved by means of low-loss micro-resonators \cite{cavity-QED}.
An alternative suggested by Andre and Lukin \cite{Andre-PRL-2002} and
first implemented in a proof-of-principle experiment by Bajcsy et al. \cite{Bajcsy-Nature-2003},
is to create spatially confined quasi-stationary pulses of light with very
low losses by means of electromagnetically induced transparency (EIT)
\cite{Harris-Physics-Today-1997,Fleischhauer-RMP-2005} in an ensemble of $\Lambda$-type
three-level atoms driven by two counter-propagating control fields.
The physical properties of stationary light were discussed in a number of
theoretical studies. It was shown that under adiabatic conditions
quasi-stationary light obeys a Schr\"{o}dinger equation with complex mass and that
inhomogeneous control fields can be used to spatially confine and compress
its wave-packet \cite{Zimmer-OptComm-2006}. The fundamental quasi-particles
of stationary light have been identified \cite{Zimmer-PRA-2008}, a transition from
a Schr\"odinger-like to a Dirac-like dynamics has been found \cite{Unanyan-2008}
and many-body phenomena with stationary-light polaritons have been discussed
theoretically such as the Tonks gas transition \cite{Chang2008} and
Bose-Einstein condensation \cite{Fleischhauer-PRL-2008}.

An essential assumption of the original model for stationary light is the
secular approximation in which spatial modulations of the ground-state coherence of the $\Lambda$-type
atoms with wavenumbers on the order of the optical fields and its harmonics are neglected.
The latter is a very good approximation in warm gases, where atomic motion
leads to a fast dephasing of fast spatial oscillations. It fails however
for cold gases or other systems where the motion is suppressed such as solids
\cite{solids} or atoms in optical lattices \cite{Masalas2004,Trotzky-2008}.
Although the problem of a secular approximation can be entirely avoided by
using a double-$\Lambda$ rather than a $\Lambda$ transition \cite{Moiseev2,Zimmer-PRA-2008},
it is interesting to consider the dynamics in a cold gas of $\Lambda$-type
atoms without the secular approximation. The earliest analysis of this case was
done by Moiseev and Ham \cite{Moiseev}. However their analysis
was limited to the case of unequal coupling field intensities, thus the
probe field was not really stationary.
In a more recent study M\o lmer and Hansen found that
without secular approximation the wave-packet of light is truly stationary
for equal strength of the control field, i.e. does not undergo a diffusive spreading, and the only
source of photon loss is the finite lifetime of the ground-state coherence
\cite{Molmer}. In this analysis radiative losses where neglected however. The result obtained in
\cite{Molmer} predicts the possibility of light trapping in EIT media with unprecendented $Q$ factors.
In the present paper we analyze stationary light in $\Lambda$-type media
without the secular approximation by taking into account the relaxation of the excited state.
We prove that the unavoidable excited-state decay limits the lifetime of the probe field in the medium.
It leads either to a
broadening of the quasi-stationary wavepacket in time or a splitting into two pulses
\cite{Xue-PRA-2008} depending on the system parameters.
The general dynamical behaviour is non-trivial,
leading e.g. to a
non-exponential decay of photons from the initial volume. We identify
parameter regimes in which the effective
loss in cold gases is slower or faster than the one in a warm gas where the
secular approximation holds.

%%%%%%%%%%%%%%%%%%%%%%%%%%%%%%%%%%%%%%%%%%%%%%%%%

\section{field equations of stationary light beyond the secular approximation%
}

%%%%%%%%%%%%%%%%%%%%%%%%%%%%%%%%%%%%%%%%%%%%%%%%%%

\begin{figure}[hbt]
    \begin{center}
   \includegraphics[width=6cm]{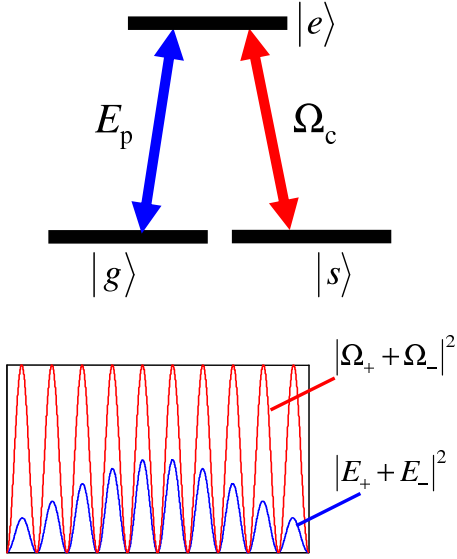}
      \caption{(color online) Schematic diagram of the three-level atomic
 system and the medium pumped by a standing wave coupling field.
 }
      \label{fig:system}
    \end{center}
\end{figure}

We here consider a medium consisting of an ensemble of non-moving three-level
atoms with a $\Lambda $ configuration shown in Fig.\ref{fig:system}. We
assume that initially some coherence is stored in the lower levels of the atomic medium,
so that when a standing
wave resonant coupling field $\Omega _{c}$ is applied, a quasi-stationary
probe field
$E$ is created. For simplicity the states $\left\vert g\right\rangle $ and $\left\vert s\right\rangle $
are assumed to be degenerate, thus the wave vectors of the
probe and the coupling fields have equal magnitude $k$.

%%%%%%%%%%%%%%%%%%%%%%%%%%%%%%%%%%%%%%%%%%%%%%%%%%%%%%%%%%%%%%%%%%%%%%%%%%
%%%%%%%%%%%%%%%%%%%%%%%%%%%%%%%%%%%%%%%%%%%%%%%%%%%%%%%%%%%%%%%%%%%%%%%%%%

The interaction Hamiltonian in rotating wave approximation reads
\begin{equation}
\hat{H}=-\hbar \sum_j\left( gE\hat \sigma _{eg}^j+\Omega _{c}\hat\sigma
_{es}^j\right) + H.c.,  \label{hamiltonian}
\end{equation}
where $E$ is the dimensionless slowly-varying complex amplitude of the probe
field, $g$ is the atom field coupling constant, $\Omega _{c}$ is the Rabi
frequency of the coupling field, and $\hat\sigma _{\mu \nu}^j=\left\vert \mu\right\rangle_{jj}
\left\langle \nu\right\vert $
are the atomic transition operators of the $j$th atom
between states $|\mu\rangle$ and $|\nu\rangle$. The atom
dynamics is governed by Langevin equations corresponding to
(\ref{hamiltonian}) and including losses from the excited state.
They can be written as
\begin{eqnarray}
\partial _{t}\hat{\sigma}_{ge} &=&-\Gamma \hat{\sigma}_{ge}+igE(\hat{\sigma}_{gg}-\hat{\sigma}_{ee})+i\Omega _{c}%
\hat{\sigma}_{gs},  \label{Langevin} \\
\partial _{t}\hat{\sigma}_{gs} &=&i\Omega _{c}^{\ast }\hat{\sigma}_{ge}+igE\hat{\sigma}_{es},
\end{eqnarray}
where $\Gamma $ is the relaxation rate of the upper level and it is
assumed that the decoherence of the lower-level transition is negligible
on the time scale of interest. In the limit of low probe-field intensities ($|gE|\ll |\Omega _{c}|$) and for an initial preparation of the ensemble in the ground state $|g\rangle$, we can set in eq.(\ref{Langevin}) $\hat{\sigma}_{gg}\approx\hat{1};\quad \hat{\sigma}_{ss}=\hat{\sigma}_{ee}=\hat{\sigma}_{se}=0$, which corresponds to the well-known pertubative linear-response limit. Since losses are included in the above
equations there should be in general Langevin noise operators \cite{Gardiner}.
The noise operators are however inconsequential since in the linear-response limit, considered here,
there is no excitation of the excited states. Thus they are neglected. We assume furthermore, that the
characteristic duration of interaction $T$ is long compared with respect to
the upper level relaxation ($\Gamma T\gg 1)$. This allows for
an adiabatic elimination of the optical coherence
$\hat\sigma _{ge}$ and equations (\ref{Langevin}) are reduced
to
\begin{align}
\Gamma \hat{\sigma}_{ge}& =igE+i\Omega _{c}\hat{\sigma}_{gs}
\label{Langevin2} \\
\partial _{t}\hat{\sigma}_{gs}& =i\Omega _{c}^{\ast }\hat{\sigma}_{ge}.
\notag
\end{align}
Differentiating the first equation with respect to time and assuming a
constant control field $\Omega_c$ yields
\begin{equation*}
\partial _{t}\hat{\sigma}_{ge}=\frac{ig}{\Gamma }\partial _{t}E-\frac{%
|\Omega _{c}|^{2}}{\Gamma }\hat{\sigma}_{ge},
\end{equation*}
which has the formal solution
\begin{eqnarray}
\hat{\sigma}_{ge}\left( z,t\right) &=&\frac{ig}{\Gamma }\int\limits_{0}^{t}\frac{\partial E}{\partial \tau}\exp \left\{ -%
\frac{|\Omega _{c}|^{2}}{\Gamma }(t-\tau )\right\} {\rm d}\tau  \notag \\
&&+i\frac{\Omega _{c}}{\Gamma }\hat{\sigma}_{gs}\left( z,0\right) \exp
\left\{ -\frac{|\Omega _{c}|^{2}}{\Gamma }t\right\}.  \label{coherence}
\end{eqnarray}
Since the coupling field is a standing wave formed by two
counterpropagating fields of equal intensity and polarization,
it can be expressed as $\Omega_{c}\left( z\right) =2\Omega \cos \left(kz\right) $, where
$\Omega $ represents the amplitude of the coupling field. The probe field
consists also of two counterpropagating components $E=E_{+}e^{ikz}+E_{-}e^{-ikz}$.
Due to the presence of $|\Omega_c(z)|^2$ in the exponents in eq.\ref{coherence},
the optical coherence and thus by virtue of eq.\ref{Langevin2} also the
ground-state coherence will develop all harmonics of $e^{\pm ikz}$. Thus we make the
ansatz
\begin{eqnarray}
\hat\sigma _{gs}\left(z, t\right) &=&\sum\limits_{n=-\infty }^{\infty
}\hat\sigma _{gs}^{(n)}\left(z, t\right) e^{iknz},  \label{Fourier} \\
\hat\sigma _{ge}\left( z,t\right) &=&\sum\limits_{n=-\infty }^{\infty
}\hat\sigma _{ge}^{(n)}\left( z,t\right) e^{iknz}.  \notag
\end{eqnarray}
The secular approximation corresponds to disregarding all
terms in $\hat{\sigma}_{gs}$ with $n\neq 0$. This is justified in a hot gas
where atomic motion washes out the fast spatial oscillations associated with
terms $\hat{\sigma}_{gs}^{(n)}$ and $n\neq 0$ \cite{Andre-PRL-2002},\cite{Zimmer-OptComm-2006}.

The propagation of the probe pulse components are governed by the Maxwell
equations for the slowly varying field amplitudes
\begin{equation}
\left( \frac{\partial }{\partial t}\pm c\frac{\partial }{\partial z}\right)
E_{\pm }=igN\hat{\sigma}_{ge}^{(\pm 1)}.  \label{Maxwell}
\end{equation}
where $\hat\sigma _{ge}^{(\pm 1)}$ are the components of atomic coherence
between levels $\left\vert g\right\rangle $ and $\left\vert e\right\rangle $ that
oscillate in space according to $e^{\pm ikz}$, and $N$ is the number density
of atoms.

If the stationary light pulse is generated from a stored spin coherence without rapidly oscillating
components, i.e. for $\sigma_{gs}^{(\pm n)}(z,t=0)=0$, for $n\ne0$
the corresponding initial conditions are
\begin{eqnarray}
\Gamma\hat\sigma_{ge}^{\pm 1}(z,0) &=& i\Omega\,\hat \sigma_{gs}(z,0),
\label{incond} \\
\hat\sigma_{ge}^{\pm n}(z,0) &=& 0,\quad\text{for}\enspace n\ne 1.  \notag
\end{eqnarray}
Using the identity $e^{A\cos z}=\sum\limits_{n=-\infty }^{\infty
}I_{\left\vert n\right\vert }\left( A\right) \cos \left( zn\right)
=\sum\limits_{n=-\infty}^{\infty} I_{\left\vert n\right\vert }\left(
A\right) \text{e}^{i z n} $,
with $I_n$ being the $n$th order modified Bessel function, we
can rewrite equation (\ref{coherence}).
\begin{eqnarray}
&&\qquad \qquad \hat{\sigma}_{ge}^{(\pm 1)}=\frac{ig}{\Gamma }%
\,\int\limits_{0}^{t}\mathrm{d}t^{\prime }\,e^{-\beta (t-t^{\prime })}\times
\label{coher2} \\
&&\times \left[ I_{0}\left( \beta (t-t^{\prime })\right) \frac{\partial
E_{\pm }}{\partial t^{\prime }}-I_{1}\left( \beta (t-t^{\prime })\right)
\frac{\partial E_{\mp }}{\partial t^{\prime }}\right]  \notag \\
&&\enspace+i\hat{\sigma}_{gs}(z,0)\Omega\mathrm{e}^{-\beta t}\left[
I_{0}\Bigl(\beta t\Bigr)-I_{1}\Bigl(\beta t\Bigr)\right]/\Gamma,  \notag
\end{eqnarray}
where we have introduced $\beta =2 \Omega^2/\Gamma$.
If we consider times $t$ which are sufficiently large, the initial
value term in eq.(\ref{coher2}) can be disregarded.

Substituting (\ref{coher2}) and (\ref{incond}) into
equations (\ref{Maxwell}) and introducing the sum and difference normal modes
$E_{s}=E_{+}+E_{-}$, $E_{d}=E_{+}-E_{-}$ yields
\begin{align}
\frac{\partial E_{s}}{\partial t}+c\frac{\partial E_{d}}{\partial z}& =-%
\frac{g^{2}N}{\Gamma }\int\limits_{0}^{t}\mathrm{d}t^\prime
f_-\Bigl(\beta(t-t^\prime)\Bigr) \frac{\partial E_s}{\partial t^{\prime }}
\label{dynamics} \\
\frac{\partial E_{d}}{\partial t}+c\frac{\partial E_{s}}{\partial z}& =-%
\frac{g^{2}N}{\Gamma }\int\limits_{0}^{t}\mathrm{d}t^\prime
f_+\Bigl(\beta(t-t^\prime)\Bigr)\frac{ \partial E_d}{\partial t^{\prime }},
\label{dynamicsd}
\end{align}
where $f_\pm (x) = e^{-x}\Bigl[I_0\bigl(x\bigr) \pm I_1\bigl(x\bigr)\Bigr]$.

% The functions $f_\pm(x)$ are shown in Fig.\ref{fig:function-pm}. One recognizes that both

% functions have a sharp peak at $x=0$ with width of order unity.

%%%%%%%%%%%%%%%%%%%%%%%%%%%%%%%%%%%%%%%%%%%%%%%%%%%%%%%%%%%%%%%%%%%%%%%%%%
% \begin{figure}[hbt]
%    \begin{center}
%   \includegraphics[width=6 cm]{/fig2.eps}
%      \caption{integral kernel of eqs.(\ref{dynamics})}
%      \label{fig:function-pm}
%    \end{center}
% \end{figure}
%%%%%%%%%%%%%%%%%%%%%%%%%%%%%%%%%%%%%%%%%%%%%%%%%%%%%%%%%%%%%%%%%%%%%%%%%%

For the following discussion it is convenient to introduce normalized
variables and parameter
\begin{equation}
\tau \equiv \frac{g^2 N}{\Gamma}\, t,\qquad \xi \equiv \frac{z}{l_{\mathrm{abs}}},\qquad a\equiv \frac{2 \Omega^2}{g^2 N}=2\cot^2\theta,
\end{equation}
where $l_{\mathrm{abs}}= c \Gamma/g^2 N$ is the resonant absorption length
of the medium in the absence of EIT. This leads to the normalized equations
\begin{eqnarray}
\partial _{\tau }E_{s}+\partial _{\xi }E_{d} &=&-\int_{0}^{\tau }\!\!\mathrm{
d}\tau ^{\prime }\,f_{-}\bigl(a(\tau -\tau ^{\prime })\bigr)\,\partial
_{\tau ^{\prime }}E_{s},  \label{eq:dynamics-s} \\
\partial _{\tau }E_{d}+\partial _{\xi }E_{s} &=&-\int_{0}^{\tau }\!\!\mathrm{
d}\tau ^{\prime }\,f_{+}\bigl(a(\tau -\tau ^{\prime })\bigr)\,\partial
_{\tau ^{\prime }}E_{d}.  \label{eq:dynamics-d}
\end{eqnarray}
%
%
%%%%%%%%%%%%%%%%%%%%%%%%%%%%%%%%%%%%%%%%%%%

\section{probe-field dynamics}

%%%%%%%%%%%%%%%%%%%%%%%%%%%%%%%%%%%%%%%%%%%%

In the following we will qualitatively discuss the probe-field dynamics
resulting from eqs.(\ref{eq:dynamics-s},\ref{eq:dynamics-d}), illustrate
the results with numerical examples and compare the field evolution with the
case of a hot atomic gas.
Eqs.(\ref{eq:dynamics-s},\ref{eq:dynamics-d}) turn into the corresponding
equations for a warm atomic gas where the secular approximation is valid, if one sets
$f_-(a (\tau-\tau^\prime))\sim (2/a)\delta(\tau-\tau^\prime)$, and $f_+(a (\tau-\tau^\prime))\to 1$
\begin{eqnarray}
\frac{\partial E_{s}}{\partial \tau}+\frac{\partial E_{d}}{\partial \xi}& =
&-\tan^2\theta \frac{\partial E_{s}}{\partial \tau}, \\
\frac{\partial E_{d}}{\partial \tau}+\frac{\partial E_{s}}{\partial \xi}& =
&-E_{d},
\end{eqnarray}
where $\tan^2\theta = g^2 N/\Omega^2$.
In this case adiabatic eliminating the fast decaying difference mode,
i.e. $E_d \simeq -\partial_\xi E_s$,
results in a diffusion equations for the sum mode
\begin{eqnarray}
\frac{\partial E_{s}}{\partial \tau} + \frac{\partial^2 E_s}{\partial \xi^2}
= 0,\qquad \\
\text{or}\qquad \frac{\partial E_{s}}{\partial t} + v_{\mathrm{gr}} l_{%
\mathrm{abs}} \frac{\partial^2 E_s}{\partial z^2} = 0,
\end{eqnarray}
where $v_{\mathrm{gr}} = c \cos^2\theta$ is the group velocity
of EIT. Associated with the diffusion is a (non-exponential) loss of
excitation with a characteristic
time scale of $T_{\mathrm{loss}}^{s}= {L^2}/{l_{\mathrm{abs}} v_{\mathrm{gr}}}$,
with $L$ being the characteristic initial confinement length of the
stationary pulse.

\begin{figure}[hbt]
    \begin{center}
   \includegraphics[width=6 cm]{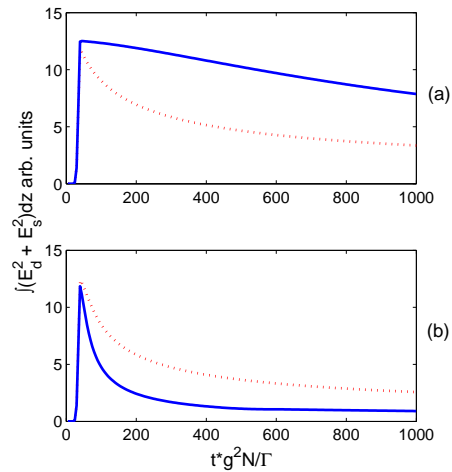}
      \caption{(color online) Decay of total field intensity
	within the spatial region from -3 $L_0$ to 3 $L_0$ for an initial spin
	excitation of spatial shape $\exp\{-z^2/L_0^2\}$, $L_0=5 l_{\mathrm{abs}}$
	in cold atomic gas (blue, solid line)\ and in inhomogeneously broadened
	(red, dotted line) media for $\tan^2\protect\theta = 100$ (top) and $\tan^2%
	\protect\theta =1$ (bottom).}
      \label{fig:decay}
    \end{center}
\end{figure}

In order to discuss the stationary-light dynamics beyond the secular
approximation we start with numerical solutions of eqs.(\ref{eq:dynamics-s},\ref%
{eq:dynamics-d}) for two characteristic cases.
In Fig.\ref{fig:decay} the decay of the total field intensity $I=\int
\mathrm{d}z (E_+^2+E_-^2)$ in the interval $\{-3 L_0,3L_0\}$ is shown after retrieval of an initial
gaussian spin excitation of spatial shape
$\exp\bigl\{-z^2/L_0^2\bigr\}$, and $L_0= 5 l_{\mathrm{abs}}$ (solid line)
for two important cases.
In the first case (top curve) $\tan^2\theta = 100$, i.e. $a=0.02$, in the
second (bottom curve) $\tan^2 \theta =1$, i.e.
$a=2$. Also shown is a comparision with the results obtained with the
secular approximation (dotted line).

%%%%%%%%%%%%%%%%%%%%%%%%%%%%%%%%%%%%%%%%%%%%%%%%%%%%%%%%%%%%%%%%%%%%%%%%%%
%%%%%%%%%%%%%%%%%%%%%%%%%%%%%%%%%%%%%%%%%%%%%%%%%%%%%%%%%%%%%%%%%%%%%%%%%%

\begin{figure}[hbt]
    \begin{center}
   \includegraphics[width=6 cm]{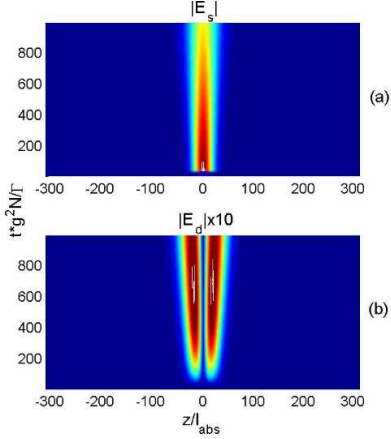}
      \caption{(color online) Spatio-temporal dynamics of sum $E_s$ (a) and
difference $E_d\times10$ (b) modes of the probe field corresponding to the top curve in
Fig.\protect\ref{fig:decay}, i.e. $\tan^2\protect\theta=100$.}
      \label{fig:field-100}
    \end{center}
\end{figure}

\begin{figure}[hbt]
    \begin{center}
   \includegraphics[width=6 cm]{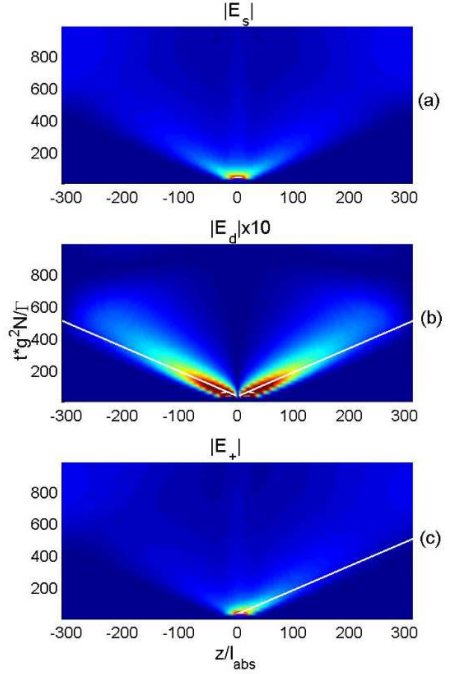}
      \caption{(color online) Spatio-temporal dynamics of sum $E_s$ (a) and
difference $E_d\times10$ (b) modes of the probe field, as well as field component $E_+$
(c), corresponding to the bottom curve in Fig.\protect\ref{fig:decay}, i.e. $%
\tan^2\protect\theta=1$. The white lines indicate the velocity $2c\cos^2%
\protect\theta/(2\cos^2\protect\theta+\sin^2\theta)$.
}
      \label{fig:field-1}
    \end{center}
\end{figure}

\noindent The time evolution of the field distributions of $E_s$ and $E_d$
for the two cases are shown in Fig.\ref{fig:field-100} ($\tan^2\theta = 100$%
) and Fig.\ref{fig:field-1} ($\tan^2\theta =1$).

%%%%%%%%%%%%%%%%%%%%%%%%%%%%%%%%%%%%%%%%%%%%%%%%%%%%%%%%%%%%%%%%%%%%%%%%%%
%%%%%%%%%%%%%%%%%%%%%%%%%%%%%%%%%%%%%%%%%%%%%%%%%%%%%%%%%%%%%%%%%%%%%%%%%%

%%%%%%%%%%%%%%%%%%%%%%%%%%%%%%%%%%%%%%%%%%%%%%%%%%%%%%%%%%%%%%%%%%%%%%%%%%
%%%%%%%%%%%%%%%%%%%%%%%%%%%%%%%%%%%%%%%%%%%%%%%%%%%%%%%%%%%%%%%%%%%%%%%%%%

\begin{figure}[hbt]
    \begin{center}
   \includegraphics[width=6 cm]{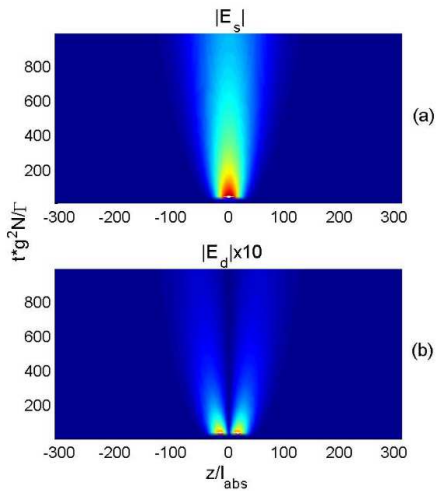}
      \caption{(color online) Dynamics of the sum $E_s$ (a) and
difference modes $E_d\times10$ (b) of the probe field in inhomogeneosly broadened medium, i.e.
under conditions that justify the secular approximation. All parameteres are
identical to the Fig.\protect\ref{fig:decay}.
}
      \label{fig:field-secular}
    \end{center}
\end{figure}

From the numerical examples several conclusions can be drawn: First of all
one recognizes that contrary to the claims in Ref.\cite{Molmer} the field
intensity decays even if the dephasing of the ground-state coherence is
neglected. The decay is caused by the relaxation of the upper state which was not taken into account in \cite{Molmer} by
restricting the discussion to the loweest order in the adiabatic expansion.
Thus stationary light in cold gases or solids does not provide a perfect
cavity. Secondly the decay of the intensity can either be slower or faster as
compared to the case with secular approximation. In the first case, i.e.
Fig.\ref{fig:field-100} the
evolution of the field distribution is very similar to the diffusive
spreading but much slower than in the secular-approximation limit shown in
Fig.\ref{fig:field-secular}. On the
other hand in the second case (see Fig.\ref{fig:field-1}), two pulse
components emerge
which propagate with the group velocity $v_{\mathrm{gr}}= c \cos^2\theta$
with some additional
loss \cite{Xue-PRA-2008}.

%%%%%%%%%%%%%%%%%%%%%%%%%%%%%%%%%%%%%%%%%%%%%%%%%%%%%%%%%%%%%%%%%%%%%%%%%%
%%%%%%%%%%%%%%%%%%%%%%%%%%%%%%%%%%%%%%%%%%%%%%%%%%%%%%%%%%%%%%%%%%%%%%%%%%

We now want to give a qualititative explanation of the different dynamics
in the two cases, which is due to the different action of the integral
kernels $f_\pm$
in eqs.(\ref{eq:dynamics-s}) and (\ref{eq:dynamics-d}). For this
it is instructive to perform a Laplace-transform of eqs.(\ref{eq:dynamics-s}) and (\ref{eq:dynamics-d}):
\begin{eqnarray}
s\widetilde{E}_{s}-E_{s}(0)+\partial _{\xi }\widetilde{E}_{d} &=&-
\widetilde{f}_{-}(s)\cdot \left( s\widetilde{E}_{s}-E_{s}(0)\right) ,
\label{eq:Es-Laplace} \\
s\widetilde{E}_{d}-E_{d}(0)+\partial _{\xi }\widetilde{E}_{s} &=&-
\widetilde{f}_{+}(s)\cdot \left( s\widetilde{E}_{d}-E_{d}(0)\right) ,
\label{eq:Ed-Laplace}
\end{eqnarray}
where the Laplace transform of $f_\pm$ reads
\begin{eqnarray*}
{\widetilde{f}}_{+}(s) &=&\frac{1}{a}\left[ \sqrt{2\frac{a}{s}+1}-1\right] ,
\\
{\widetilde{f}}_{-}(s) &=&\frac{1}{a}\left[ -\frac{1}{\sqrt{2\frac{a}{s}+1}}%
+1\right] .
\end{eqnarray*}
This yields
\begin{eqnarray}
s\widetilde E_s-E_s(0) &=& -\frac{\partial_\xi \widetilde E_d }{1+{%
\widetilde f}_-(s)}, \\
s\widetilde E_d-E_d(0) &=& - \frac{\partial_\xi \widetilde E_s }{1+{%
\widetilde f}_+(s)}.
\end{eqnarray}
%

%*********************************************************************

\subsubsection{limit of small $a$}

%********************************************************************

Assuming that $a$ is small, a series expansion of $1/(1+{\widetilde f}_\pm(s))$ yields
\begin{equation*}
\frac{1}{1+{\widetilde{f}}_{\pm }(s)}\approx \frac{s}{s+1}
\end{equation*}
For the slow time evolution, i.e. for (physical) times large compared to $%
(g^2 N/\Gamma)^{-1}$ only values of $s\ll 1$ are relevant, and thus the right hand side can be
replaced by $s$. Substituting this into eqs.(\ref{eq:Es-Laplace}) and
(\ref{eq:Ed-Laplace}) one arrives at
\begin{eqnarray*}
\partial _{\xi }^{2}E_{s}-E_{s} &=&-\left( E_{s}(\xi ,0)+\partial _{\xi
}E_{d}(\xi ,0)\right) , \\
\partial _{\xi }^{2}E_{d}-E_{d} &=&-\left( E_{d}(\xi ,0)+\partial _{\xi
}E_{s}(\xi ,0)\right),
\end{eqnarray*}
which describes truly stationary wave packets that decay exponentially with
increasing distance. That there is no dynamics is of course due to
the fact that only the leading order term in the expansion of $1/(1+{\widetilde f}_\pm)$
was taken into account.

%*********************************************************************

\subsubsection{limit of large $a$}

%********************************************************************

A qualitative explanation of the opposite case can be found by considering
the limit of large $a$. To properly analyze this case one has to take into account that also the
most relevant Laplace frequency $s$ increases when $a$ becomes large. In fact the numerical
data suggest that the ratio $a/s$, with $s$ being the most relevant Laplace frequency
approaches a constant. Thus in
this case one has
\begin{eqnarray*}
\frac{1}{1+{\widetilde{f}}_{+}(s)} &\approx &\frac{a}{a-1+x} \\
\frac{1}{1+{\widetilde{f}}_{-}(s)} &\approx &\frac{a}{a+1-1/x},
\end{eqnarray*}
where $x=\sqrt{2\dfrac{a}{s}+1}$ is well approximated by a constant. This leads to the approximate
equations
\begin{eqnarray*}
\partial _{\tau }E_{s}+\frac{a}{a+1-1/x}\partial _{\xi }E_{d} &=&0, \\
\partial _{\tau }E_{d}+\frac{a}{a-1+x}\partial _{\xi }E_{s} &=&0.
\end{eqnarray*}

In the limit of large $a\,$\ one arrives at wave equations for the forward
and backward components
\begin{equation*}
\partial _{\tau }^{2}E_{\pm }{\pm}\frac{a^2}{(a+1-1/x)(a-1+x)}
\partial _{\xi }^{2}E_{\pm }=0,
\end{equation*}
which reads in physical time and space:
\begin{equation*}
\partial _{t}^{2}E_{\pm }{\pm}\frac{c^{2} a^2}{(a+1-1/x)(a-1+x)}\partial _{z}^{2}E_{\pm }=0.
\end{equation*}
Thus the envelope of $E_{\pm }$ evolves freely
This explains the splitting
of the stationary light wavepacket into two
components each of which propagating with the modified group velocity
$2c \cos^2\theta/(2\cos^2\theta +f\sin^2\theta)$, with $f=(x-1/x)/4$.
Noting that the most
relevant Laplace frequency for the example in Fig.\ref{fig:field-1} leads to a value of $f$ on the order
of unity we find reasonable agreement with the numerical results.
%%%%%%%%%%%%%%%%%%%%%%%%%%%%%%%%%%%%%%%%%%%%%%%%%%%%%%%%%%%%%%%%%%%%%%%

\section{summary}

%%%%%%%%%%%%%%%%%%%%%%%%%%%%%%%%%%%%%%%%%%%%%%%%%%%%%%%%%%%%%%%%%%%%%%%%%5

We considered the dynamics of stationary light in a standing medium without
secular approximation and derived equations describing the evolution of
the sum and difference modes of the pulse. A numerical as well as approximate
analytical solution showed  that for small coupling field intensities the probe field
spreading is slower than in the secular approximation but in contrast to the
results of \cite{Molmer} non-zero.
In the opposite limit of strong
coupling the probe pulse splits into two counterpropagating components.

G.N. acknowledges support by the Alexander von Humboldt Foundation.


\begin{thebibliography}{99}
\bibitem{cavity-QED} J.H. Kimble, Phys. Scr. T \textbf{76}, 127 (1998). J.M.
Raimond, M. Brune, S. Haroche, Rev. Mod. Phys. \textbf{73}, 565 (2001). H.
Walther, B.T.H. Varcoe, B.G. Englert and Th. Becker, Rep. Prog. Phys.
\textbf{69}, 1325 (2006).

\bibitem{Andre-PRL-2002} A. Andre and M.D. Lukin, Phys. Rev. Lett. \textbf{89%
}, 143602 (2002); A. Andre, M. Bajcsy, A.S. Zibrov, and M.D. Lukin, Phys. Rev. Lett. \textbf{94}, 063902 (2005).

\bibitem{Bajcsy-Nature-2003} M. Bajcsy, \ A. S. Zibrov, M.D. Lukin, \ Nature
(London), \textbf{426}, 638 (2003).

\bibitem{Harris-Physics-Today-1997} S. E. Harris, Physics Today, \textbf{50}, Nr.7, 36 (1997).

\bibitem{Fleischhauer-RMP-2005} M. Fleischhauer, A. Imamoglu, and J. P. Marangos, Rev. Mod. Phys.
\textbf{77}, 663 (2005).

\bibitem{Zimmer-OptComm-2006} F. E. Zimmer, A. Andre, M. D. Lukin, and M.
Fleischhauer, Opt. Comm. \textbf{264}, 441 (2006)

\bibitem{Zimmer-PRA-2008} F. E. Zimmer, J. Otterbach, R. G. Unanyan, B. W.
Shore and M. Fleischhauer, Phys. Rev. A \textbf{77}, 063823 (2008); Y. D. Chong, and M. Soljacic, Phys. Rev. A \textbf{77}, 013823 (2008).

\bibitem{Unanyan-2008} J. Otterbach, R.G. Unanyan and M. Fleischhauer,
Phys. Rev. Lett. \textbf{102}, 063602 (2009).

\bibitem{Chang2008} D. E. Chang, V. Gritsev, G. Morigi, V. Vuleti, M. D.
Lukin, E. A. Demler, Nature Physics \textbf{4}, 884 (2008).

\bibitem{Fleischhauer-PRL-2008} M. Fleischhauer, J. Otterbach, and R.G.
Unanyan, Phys. Rev. Lett. \textbf{101}, 163601 (2008).

\bibitem{solids} A. V. Turukhin, V. S. Sudarshanam, M. S. Shahriar, J. A.
Musser, B. S. Ham, and P. R. Hemmer, Phys. Rev. Lett. \textbf{88}, 023602
(2001).

\bibitem{Masalas2004} M. Masalas and M. Fleischhauer, Phys. Rev. A \textbf{69}, 061801(R) (2004).

\bibitem{Trotzky-2008} F. Gerbier, S. Trotzky, S. Folling, U. Schnorrberger,
J. D. Thompson, A. Widera, I. Bloch, L. Pollet, M. Troyer, B.
Capogrosso-Sansone, N. V. Prokof'ev, B. V. Svistunov, Phys. Rev. Lett.
\textbf{101}, 155303 (2008).

\bibitem{Moiseev2} S.A. Moiseev and B.S. Ham, Phys. Rev. A \textbf{73},
033812 (2006).

\bibitem{Moiseev} S.A. Moiseev and B.S. Ham, Phys. Rev. A \textbf{71},
053802 (2005).

\bibitem{Molmer} K.R. Hansen and K. M\o lmer, Phys. Rev. A \textbf{75},
065804 (2007). K.R. Hansen and K. M\o lmer, Phys. Rev. A \textbf{75}, 053802
(2007).

\bibitem{Xue-PRA-2008} Yan Xue and B. S. Ham, Phys. Rev. A \textbf{78},
053830 (2008).

\bibitem{Gardiner} Quantum Noise, C.W. Gardiner, P. Zoller, (Springer,
Berlin, 2000).
\end{thebibliography}
\end{document}